# Geometry-Optimized Complex-Domain error-diffusion encoding for Fourier Single-Pixel Imaging

Chongwu Shao,[1] Yue Cao,[1] Wei Zhang,[1] Xiaopeng-Jin,[1] Yingran Shen,[1] Shijian Li*,[2,3] and Xu-Ri Yao*[1,4]

1) *Center for Quantum Technology Research and Key Laboratory of Advanced Optoelectronic Quantum Architecture and Measurements, School of Physics, Beijing Institute of Technology, Beijing 100081, China*

2) *College of Photonics and Optical Engineering, Aerospace Information Technology University, Jinan 250299, China*

3) *Shandong Key Laboratory of Intelligent Photonic Transmission and Sensing, Jinan 250299, China*

4) *State Key Laboratory of Chips and Systems for Advanced Light Field Display, School of Physics, Beijing Institute of Technology, Beijing 100081, China*

(*Electronic mail: yaoxuri@bit.edu.cn)

(*Electronic mail: lishijian@aitech.edu.cn)

This work proposes a geometry-optimized complex-domain error-diffusion encoding method for Fourier single-pixel imaging. Instead of independently binarizing multiple grayscale phase-shifting patterns, the proposed method directly represents each complex-valued Fourier basis pattern using $K$ ($K \geq 3$) weighted binary patterns while diffusing the residual error in the complex domain. A geometric interpretation is further established, revealing that the encoding process can be viewed as approximating the Fourier-basis unit circle by a regular polygon in the complex plane. Based on this geometric interpretation, practical optimization strategies are developed for $K = 3$, $K = 4$, and $K = 7$. Both numerical simulations and real-object experiments demonstrate consistently superior reconstruction quality compared with conventional phase-shifting dithering.

Single-pixel imaging (SPI)[1–3] reconstructs images using a single-pixel detector and a spatial light modulator (SLM), providing high sensitivity and broad spectral applicability. Digital micromirror devices (DMDs), with binary refresh rates exceeding 22 kHz, are among the most widely used SLMs in SPI. Fourier-SPI[4–7] directly measures scene Fourier coefficients using Fourier basis patterns and reconstructs images through the inverse Fourier transform, enabling high-quality imaging without computationally intensive iterative algorithms.

However, there is a fundamental mismatch between the continuous complex-valued Fourier basis patterns and the binary modulation capability of the DMD. Conventionally, complex-valued Fourier basis patterns are implemented using an $N$-step phase-shifting scheme, in which each Fourier coefficient is recovered from multiple grayscale patterns with different phase offsets. These grayscale patterns are then converted into binary patterns for DMD projection. Existing binary implementations can generally be divided into two categories: spatial upsampling[8] and signal dithering[9]. Spatial upsampling with error diffusion reduces quantization noise at the expense of spatial resolution, whereas signal dithering encodes each grayscale pattern into multiple binary patterns, sacrificing temporal resolution. Since SPI inherently relies on numerous sequential measurements, spatial dithering has become the dominant solution, while recent spatiotemporal encoding[10,11] and intermediate-bit modulation[12] aim to reduce the number of binary patterns required for signal dithering. Error-diffusion performance has also been improved through serpentine Sierra-Lite diffusion[13], multiple scanning paths[14], positive-negative dithering[15] and frequency-dependent optimization of the diffusion parameters guided by reconstruction quality [16,17].

Nevertheless, binarization inevitably introduces quantization noise, while error diffusion shifts it toward high frequencies, potentially interfering significantly with high-frequency Fourier components. Since target spectra vary, no single dithering strategy is universally optimal. Despite their methodological differences, existing methods optimize individual grayscale phase-shifting patterns independently. Importantly, image reconstruction in Fourier-SPI depends on the fidelity of the synthesized complex-valued Fourier basis patterns rather than on the accuracy of any individual phase-shifting pattern, which limits the achievable performance of existing binary implementation methods.

Motivated by this observation, we propose a geometry-optimized complex-domain error-diffusion framework for Fourier-SPI. Unlike conventional methods that optimize each grayscale pattern separately, the proposed method directly encodes the desired complex-valued Fourier basis pattern using $K$ weighted binary patterns and performs error diffusion entirely in the complex domain. The method is applicable to any integer $K \geq 3$. In general, a larger $K$ provides higher reconstruction quality, allowing a flexible trade-off between temporal resolution and reconstruction quality. We further establish a geometric interpretation in which the coding process can be viewed as approximating the unit circle by a regular polygon on the complex plane. Based on this geometric model, we further optimize the performance of the two practical configurations, $K = 3$ and $K = 4$, and improve the quality of reconstruction for the case $K = 7$. Numerical simulations and real-object experiments validate the proposed framework and demonstrate its superiority over conventional phase-shifting-based binary implementations in Fourier-SPI.

Each pixel of the unit-modulus complex Fourier basis pat-

tern $P$ is expressed as

$$P(x,y) = \exp\left[-j2\pi\left(f_x x + f_y y\right)\right], \tag{1}$$

where $f_x$ and $f_y$ denote the normalized horizontal and vertical spatial frequencies, respectively. Since the Fourier basis has unit modulus, all possible values of $P(x,y)$ lie on the unit circle in the complex plane.

The proposed coding method aims to approximates the unit-modulus complex pattern $P(x,y)$ using $K$ binary patterns $B_k$, where $K \geq 3$. Let $b_k(x,y) \in \{0, 1\}$ denote the pixel value of the $k$th binary pattern. Under the full-circle encoding scheme, each pattern is assigned a complex weight

$$w_k = \exp\left(j\frac{2\pi k}{K}\right), \qquad k = 0,1,\ldots,K-1. \tag{2}$$

Thus, the Fourier basis pattern $P$ is represented as $\Sigma_k w_k B_k$. The complex value encoded by the $K$ binary values at pixel $(x,y)$ is

$$z(x,y) = \sum_{k=0}^{K-1} b_k(x,y) w_k = \sum_{k=0}^{K-1} b_k(x,y)\exp\left(j\frac{2\pi k}{K}\right). \tag{3}$$

Since each $b_k(x,y)$ is binary, all possible combinations generate a finite set of candidate points in the complex plane.

For odd values of $K$, the outermost candidate points form the vertices of a regular $2K$-gon. For even values of even $K$, they consist of the vertices of a regular $K$-gon together with the midpoints of its edges. The corresponding binary encodings comprise two groups of cyclic consecutive-one codewords with run lengths $\lfloor (K-1)/2 \rfloor$ and $\lfloor (K+1)/2 \rfloor$, respectively. Each group contains $K$ codewords obtained by cyclically shifting the starting index. Specifically, the $k$-th bit of a codeword is defined as

$$b_k^{(s,L)} = \begin{cases} 1, & \langle k-s \rangle_K \in \{0,1,\ldots,L-1\}, \\ 0, & \text{otherwise}, \end{cases} \tag{4}$$

where $\langle k-s \rangle_K$ denotes the residue of $k-s$ modulo $K$, $s = 0, 1, \ldots, K-1$ specifies the starting index of the consecutive-one run, and $L \in \{\lfloor (K-1)/2 \rfloor, \lfloor (K+1)/2 \rfloor\}$ denotes its length. The corresponding normalization radius $R$ is

$$R = \left| \sum_{k=0}^{\lfloor (K-1)/2 \rfloor} w_k \right| = \begin{cases} \dfrac{1}{2\sin(\pi/2K)}, & K \text{ is odd}, \\ \dfrac{1}{\sin(\pi/K)}, & K \text{ is even}. \end{cases} \tag{5}$$

The selected polygon is naturally centered at the origin. To match the unit-modulus Fourier basis, radius normalization is applied:

$$\tilde{z}(x,y) = \frac{z(x,y)}{R}. \tag{6}$$

In the proposed method, error-diffusion is performed directly in the complex domain. Similar to color error diffusion[18,19], a single complex-valued Fourier basis pattern is encoded into $K$ binary patterns. At every pixel, the Fourier pattern value quantized to its nearest candidate value among the $2K$ normalized candidates $\tilde{z}$, and the corresponding binary codeword determines the pixel values assigned to the $K$ patterns. The resulting complex quantization error is then propagated to neighboring unprocessed pixels using the Floyd-Steinberg kernel[20] with a serpentine scan path[21] (see supplementary note 1).

After obtaining the $K$ binary patterns $B_k$, the measurements acquired by the SPI are given by

$$S_k = \sum_{x,y} I(x,y) B_k(x,y), \tag{7}$$

where $I(x,y)$ denotes the target image. The corresponding Fourier coefficient is recovered by

$$\hat{F} = \frac{\sum_{k=0}^{K-1} S_k \exp\left(j\frac{2\pi k}{K}\right)}{R}. \tag{8}$$

Finally, the reconstructed image $\hat{I}(x,y)$ is reconstructed through the discrete inverse Fourier transform on the sampled frequency set $\Omega$:

$$\hat{I}(x,y) = \mathrm{Re}\{\sum_{\Omega} \hat{F}(f_x,f_y)\exp[j2\pi(f_x x + f_y y)]\}. \tag{9}$$

Unlike independent pixel-wise quantization, error diffusion achieves accurate representation through local spatial averaging[22]. Since the Fourier pattern varies continuously (spatial averaging of adjacent candidates $\tilde{z}$), the effective representation of the dithered binary patterns naturally extends to the regular polygonal boundary, rather than being limited to its vertices. Therefore, complex-domain error-diffusion encoding can be abstracted as a geometric model in which a regular polygon approximates the unit circle. This geometric interpretation naturally suggests two complementary optimization strategies. The first strategy improves the geometric approximation by increasing the number of polygon vertices. The second strategy keeps K fixed while geometrically adjusting the target Fourier basis so that it better matches the effective representable region of the binary patterns. This strategy is particularly beneficial for small values of $K$, where the geometric mismatch is most pronounced.

The first strategy is to increase $K$, allowing the regular polygon to more closely approximate the circumscribed unit circle and thereby provide a finer representation. In principle, the proposed method is applicable to any $K \geq 3$, although the corresponding geometric models differ for odd and even values of $K$. As shown in Fig. 1(a), for odd $K$,as $K$ increases, the approximable range encoded by the 0/1 binary patterns (the regular polygon) naturally approaches the value range of the Fourier basis pattern (the unit circle).

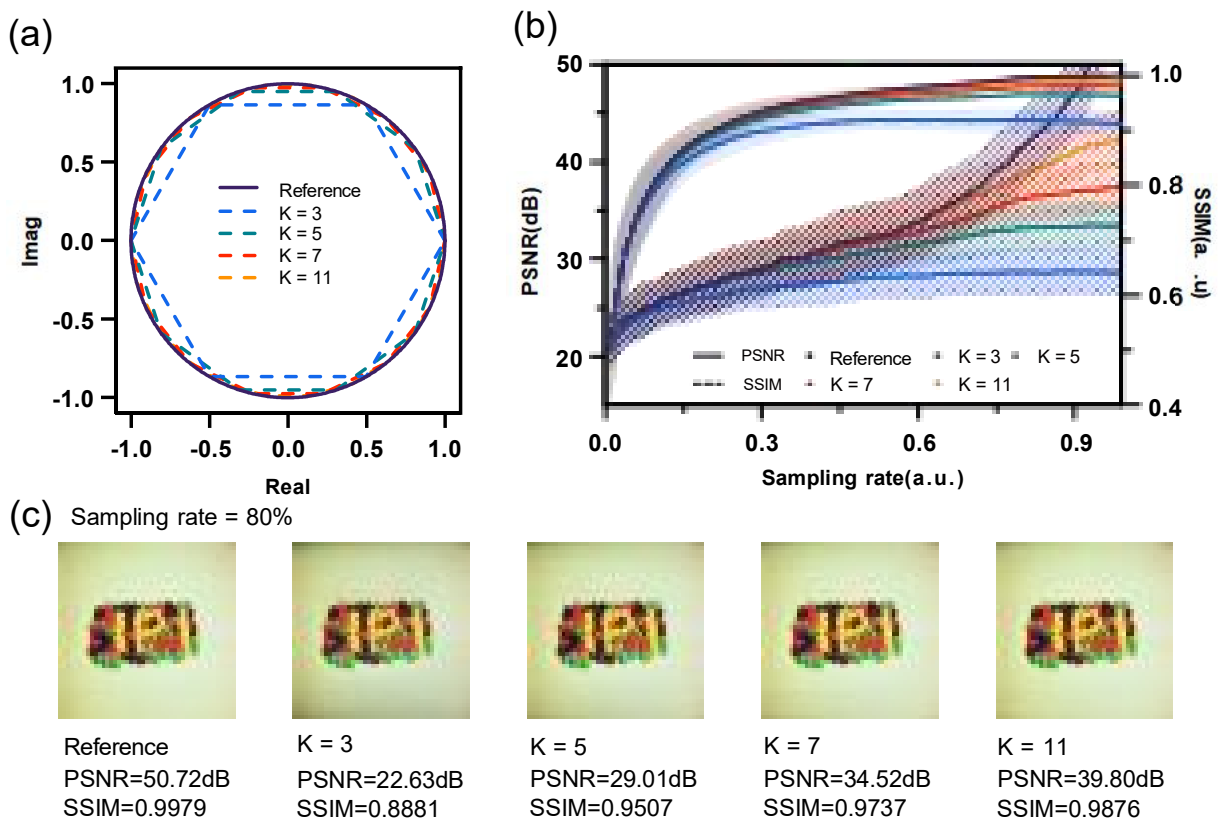

FIG. 1. Geometric interpretation and simulation validation of the Complex-domain error-diffusion method for $K = 3, 5, 7, 11$. (a) Schematic illustration of the geometric model for different values of $K$. The polygons represent the approximable ranges encoded by the $K$ binary patterns, while the unit circle represents the value range of the Fourier basis pattern. (b) PSNR and SSIM comparisons of complex-domain-error-diffusion-based Fourier-SPI reconstruction with different values of $K$ on the Kodak24 dataset. (c) Fourier-SPI reconstruction results of the Jelly Beans image using complex-domain error-diffusion with different values of$K$. All simulation results shown hereafter were obtained using $256 \times 256$ images, Floyd–Steinberg error diffusion with serpentine scanning, no upsampling, and reference reconstructions from the original Fourier basis patterns.

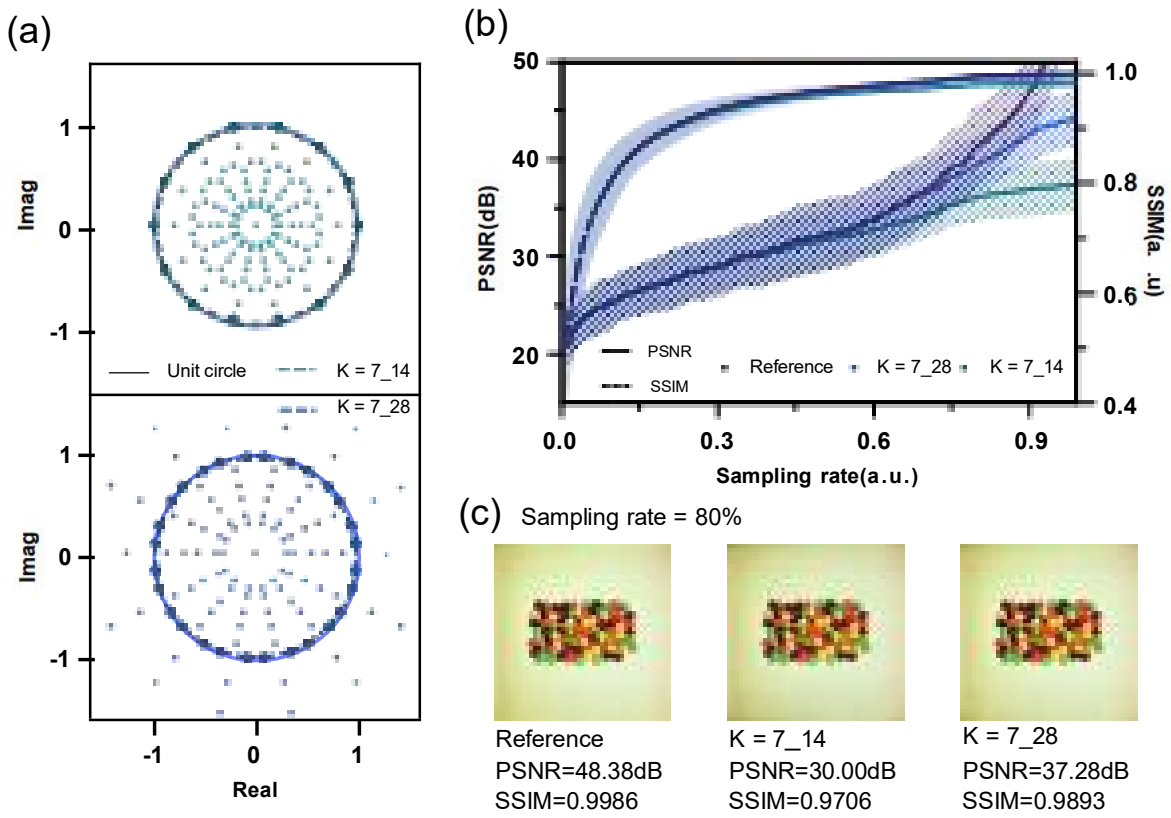

FIG. 2. Geometric model and simulation results for 14-point and 28-point encoding schemes in the $K = 7$ case. (a) Geometric models of the 14-point and 28-point encoding schemes. The 28-point encoding scheme is constructed from the cyclic codewords generated by $\{1101\}$ and $\{1011\}$, together with their complementary codewords, resulting in 28 candidate points. (b) PSNR and SSIM comparisons of the 14-point and 28-point encoding schemes on the Kodak24 dataset. (c) Fourier-SPI reconstruction results of the Jelly Beans image using the 14-point and 28-point encoding schemes.

The simulation results in Fig. 1(b) and (c) validate the geometric interpretation, showing that reconstruction quality consistently improves as the polygon more closely approximates the Fourier-basis unit circle. As $K$ increases, the 0/1 patterns provide a finer approximation to the Fourier basis patterns, thereby improving the reconstruction quality of Fourier-SPI. A similar trend is observed for even values of $K$ (see supplementary note 2). However, due to their different polygonal geometric model, odd $K$ generally outperforms comparable even $K$. In practice, the choice of $K$ requires a trade-off between the reconstruction quality and temporal resolution of SPI. Furthermore, converting a Fourier basis pattern into multiple 0/1 patterns inevitably introduces quantization errors, which can cause pronounced distortions in the sampled high-frequency components in Fourier-SPI. Therefore, in practical applications, the sampling rate should be properly controlled according to the value of $K$ to obtain more robust reconstruction results. Accordingly, all reconstruction results presented in this work are obtained at an 80% sampling rate.

For sufficiently large values of $K$, the discrete distribution of $z$ in the complex plane may contain a larger number of candidate points on certain radial layers. This observation indicates that, for a given $K$, the performance can be further improved by selecting an appropriate radial layer of candidate points. The first practically useful example appears at $K = 7$. As shown in Fig. 2(a), 28 candidate points lie on the radial layer with optimal radius $\sqrt{2}$, enabling a 28-point encoding scheme that provides a finer angular quantization without increasing the number of projected binary patterns.

Figure 2(b) and (c) show the performance improvement brought by the 28-point encoding, especially at high sampling rates. The optimized point-selection strategy reduces the quantization error introduced by binarization and mitigates high-frequency distortion, allowing Fourier-SPI to further improve its performance upper bound at high sampling rates. More generally, the optimal normalized radius $R$ exhibits a stable scaling trend as $K$ increases (see Supplementary Note 3).

The second strategy is to apply a mild geometric refinement to the circumscribed unit circle (Fourier basis patterns) for a fixed $K$, so that the resulting contour is better matched to the regular polygon. Geometric model for the practically important $K = 3$ case, the dithered binary patterns are effectively represented by a regular hexagon, while the Fourier basis pattern values lie on its circumscribed unit circle. As shown in Fig. 3(a), we consider two geometric adjustment operations: symmetric scaling of the circle and asymmetric compression into an ellipse.

The first operation is symmetric and uniformly shrinks the target unit circle by introducing a scaling factor $\alpha$. In this case, $\alpha$ is chosen as the midpoint between the circumradius and the inradius of the regular polygon. Accordingly, the modified Fourier pattern $P'(x,y)$ and the correction applied to the corresponding decoded coefficient $\widehat{F}'$ are given by

$$P'(x,y) = \alpha P(x,y), \qquad \widehat{F} = \frac{\widehat{F}'}{\alpha}, \tag{10}$$

where $\alpha = (2+\sqrt{3})/4$.

The second operation is asymmetric and compresses the target unit circle along the imaginary axis by introducing a compression factor $\beta$. In this case, $\beta$ is chosen such that the transformed contour is tangent to the regular hexagon in the imaginary-axis direction, while the original unit radius is retained along the real-axis direction. Accordingly, $P'(x,y)$ and the correction applied to $\widehat{F}'$ are expressed as

$$P'(x,y) = \mathrm{Re}(P(x,y)) + j\beta\,\mathrm{Im}(P(x,y)),$$
$$\widehat{F} = \mathrm{Re}\left(\widehat{F}'\right) + j\frac{\mathrm{Im}\left(\widehat{F}'\right)}{\beta}, \tag{11}$$

where $\beta = \sqrt{3}/2$.

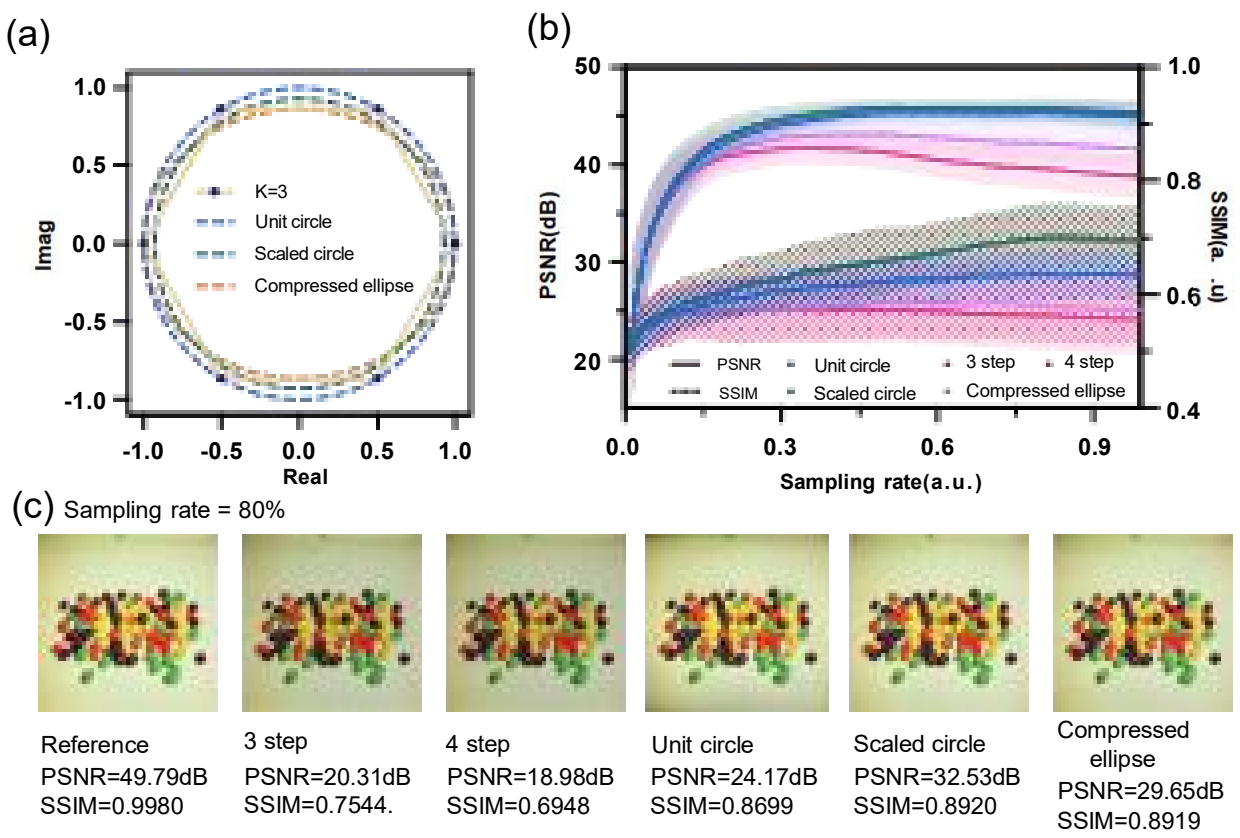

FIG. 3. Geometric model and numerical simulation of the optimization strategies for refining the Fourier-basis unit circle in the $K = 3$ case. (a) Schematic illustration of the two geometric adjustment strategies for the $K = 3$ case. (b) PSNR and SSIM comparisons on the Kodak24 dataset for unoptimized and optimized complex-domain error-diffusion in the $K = 3$ case, with conventional three-step and four-step phase-shifting methods used as baselines. (c) Fourier-SPI reconstruction results of the Jelly Beans image obtained using conventional three-step and four-step phase shifting, and unoptimized and optimized complex-domain error-diffusion for $K = 3$.

As shown in Fig. 3(b) and (c), the proposed complex-domain error-diffusion method already outperforms the conventional phase-shifting dithering method. This improvement arises because the proposed method directly approximates the complex Fourier basis patterns by diffusing residual errors in the complex domain, thereby reducing quantization error and enabling more accurate recovery of the Fourier coefficients. Building on this baseline method, the proposed geometric refinement strategies further improve the reconstruction performance by better matching the effective representation of the binary patterns to the target complex Fourier patterns.

According to the geometric model, the proposed refinement is expected to yield more pronounced improvements for smaller $K$. This is because the larger geometric mismatch provides greater room for adjustment, while the relatively weaker baseline performance makes the resulting gain more practically valuable. Therefore, we adopt the more general symmetric geometric refinement strategy and apply it to the practically important cases of $K = 3$ and $K = 4$, using scaling factors $\alpha_{K=3} = (2 + \sqrt{3})/4$ and $\alpha_{K=4} = (2 + \sqrt{2})/4$, respectively.

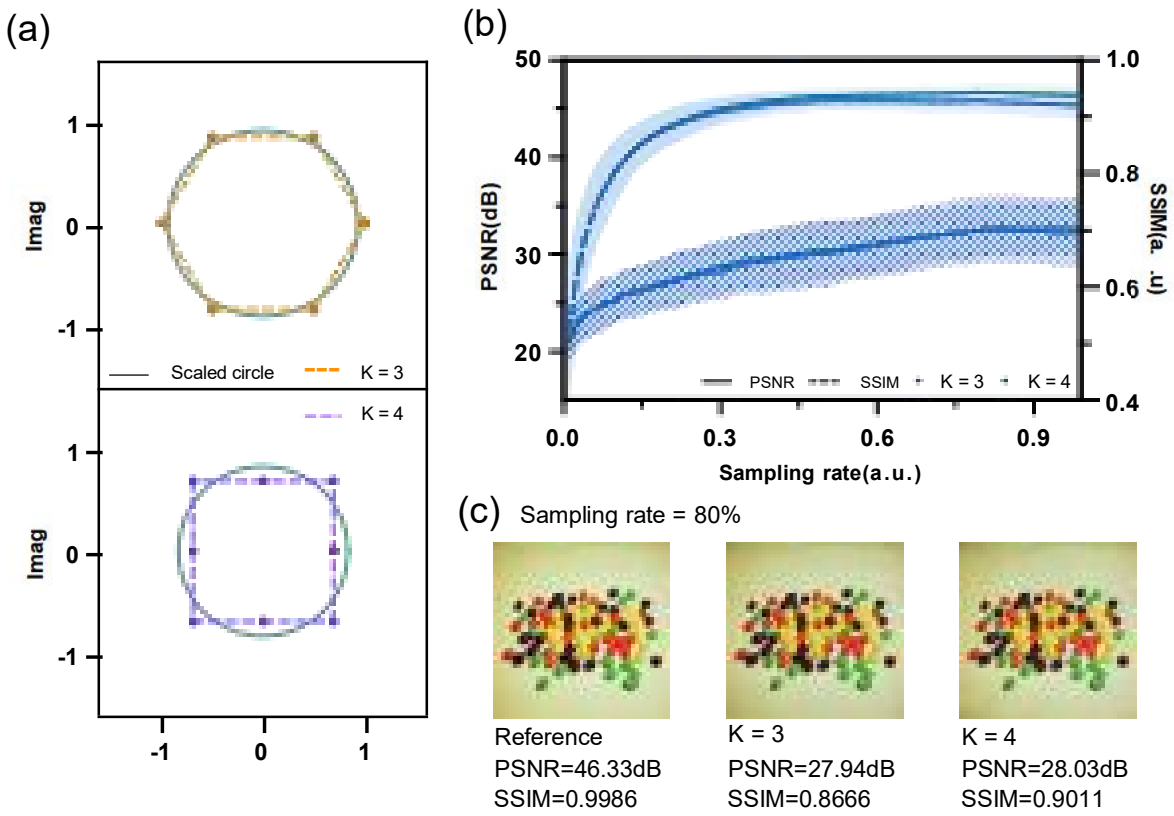

FIG. 4. Geometric model and performance evaluation of the complex-domain error-diffusion method for $K = 3$ and $K = 4$. (a) Geometric models for $K = 3$ and $K = 4$ with scaled-circle refinement. (b) PSNR and SSIM comparisons on the Kodak24 dataset for the $K = 3$ and $K = 4$ cases. (c) Fourier-SPI reconstruction results of the Jelly Beans image for the $K = 3$ and $K = 4$ cases.

When $K = 4$, the 8 coding candidate points effectively form only a quadrilateral in the complex plane, as shown in Fig. 4(a). Even after geometric refinement, their approximation to the unit circle remains weaker than that in the $K = 3$ case. As shown in Fig. 4(b) and (c), the $K = 4$ provides only a slight performance improvement over $K = 3$, and the two exhibit comparable reconstruction performance. These results suggest that more candidate points do not necessarily lead to higher reconstruction quality, and geometric matching also plays a key role. Further numerical simulations confirm the robustness and practical applicability of these scaling factors (see Supplementary Note 4).

Taken together, the simulation results confirm the effectiveness of the proposed complex-domain error-diffusion encoding method and provide practical guidance for selecting $K$ in Fourier-SPI. Considering the trade-off between reconstruction quality and temporal resolution, the symmetrically optimized $K = 3$ and $K = 4$ might be the most practical choice, while the 28-point encoding scheme for $K = 7$ provides a favorable option when higher reconstruction quality is desired.

To further validate the proposed method in practice, we conduct real-object Fourier-SPI experiments using two drawn color Zelda illustrations, with conventional three-step and four-step phase-shifting methods serving as performance baselines. The experiment employs a passive SPI optical configuration (see Supplementary Note 5).

Since ground-truth images are unavailable in the real experiments, reconstruction quality is evaluated using two customized no-reference metrics: the scene-specific natural image quality evaluator (SS-NIQE) and the wavelet-based median noise index (WMNI). SS-NIQE follows the NIQE framework[23] and measures the statistical deviation between the reconstructed image and the original scene, while WMNI estimates local high-frequency quantization noise in a smooth region using a one-level wavelet decomposition and median absolute deviation[24,25] (see Supplementary Note 6). Lower values indicate better reconstruction quality for both metrics.

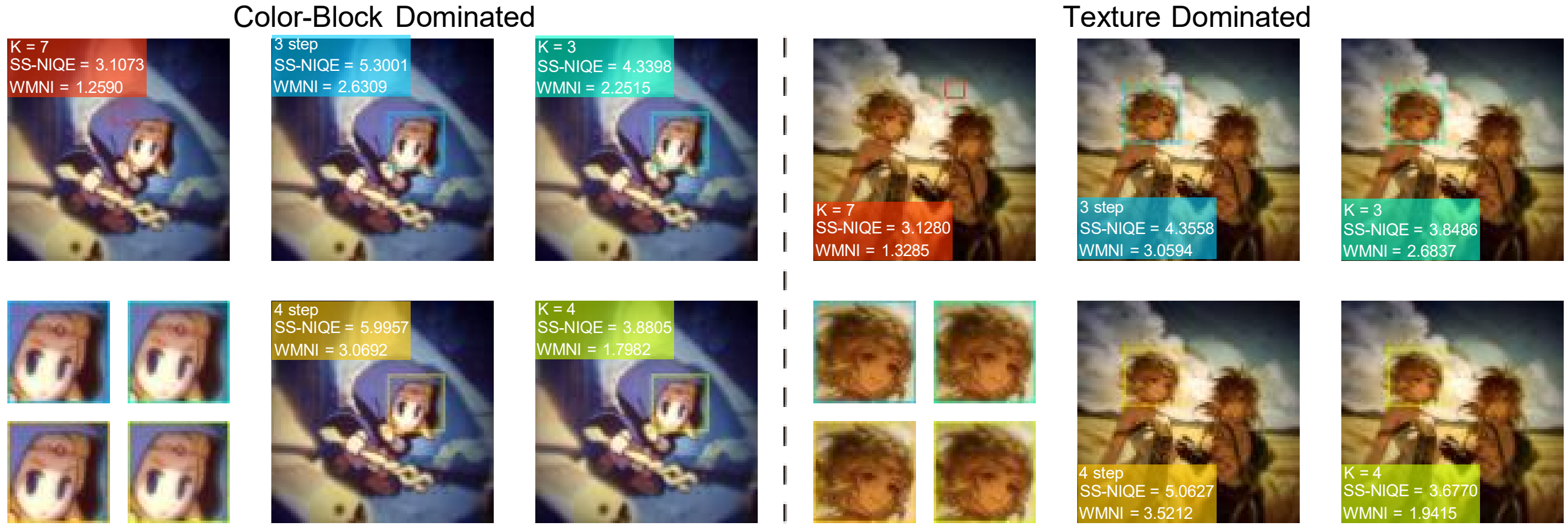

FIG. 5. Experimental validation of the proposed geometry-optimized complex-domain error-diffusion framework using real-object Fourier-SPI. Reconstructions obtained using conventional three-step and four-step phase shifting, complex-domain error-diffusion with $K = 3$ and $K = 4$, and the 28-point encoding scheme with $K = 7$. Reconstruction quality is evaluated using SS-NIQE and WMNI, with WMNI calculated over the smooth region indicated by the red box. All are acquired at an image resolution of $256 \times 256$, an 80% sampling rate, using Floyd-Steinberg error diffusion with a serpentine scanning path, and withou upsampling.

As illustrated in Fig. 5, the proposed complex-domain error-diffusion encoding method produces smoother color regions, sharper edges, and less quantization noise than conventional phase-shifting dithering. These improvements are consistently observed for both the color-dominated and texture-rich targets. The lower SS-NIQE scores indicate that the reconstructions are statistically more consistent with the predefined experimental scene, while the lower WMNI values in the smooth region confirm that the reduction in quantization noise remains detectable even in the presence of real-world experimental noise. The advantage is particularly pronounced in smooth color regions, where binary quantization artifacts are most visually apparent. These results are in good agreement with the theoretical analysis, demonstrating the effectiveness of complex-domain error diffusion in reducing binary quantization errors. The consistent performance of the proposed method under different measurement budgets is further verified by real-experiment reconstructions at different sampling rates (see Supplementary Note 7).

In summary, we propose a complex-domain error-diffusion encoding method for Fourier-SPI that directly encodes complex-valued Fourier basis pattern into $K$ binary patterns with assigned complex weights, thereby reducing quantization errors without upsampling. We further established a geometric model that interprets the coding process as a polygonal approximation to the unit circle, based on which we developed three practically relevant schemes: geometry-based optimizations for $K = 3$ and $K = 4$, and a 28-point encoding scheme for $K = 7$. Numerical simulations and real-object experiments validated the improved reconstruction performance of the proposed method over conventional phase-shifting-based dithering. More importantly, the proposed complex-domain error-diffusion encoding framework is compatible with existing error-diffusion optimization techniques for Fourier-SPI, allowing their further integration to achieve additional improvements in reconstruction performance.

See Supplementary Material for numerical and theoretical analyses of different $K$ values, scaled-circle optimization, experimental configurations, no-reference metrics, and additional real-experiment results at different sampling rates.

This work was supported by National Natural Science Foundation of China (Grant No.12504337) and Beijing Institute of Technology Research Fund Program for Young Scholars (202122012).

## AUTHOR DECLARATIONS

### Conflict of Interest

The authors declare no conflict of interest to disclose.

### Author Contributions

C.S. and S.L. conceived the research idea and developed the theoretical analysis. C.S. and Y.C. performed the numerical simulations. C.S., Y.C., X.Y., S.L., and W.Z. carried out the experiments. C.S., Y.C., X.Y., S.L., W.Z., X.J., and Y.S. discussed the results. C.S. and Y.C. prepared the manuscript, with revisions from all authors. All authors reviewed and approved the final manuscript.

## DATA AVAILABILITY STATEMENT

The raw data and code supporting the findings of this study are available from the corresponding authors upon reasonable request.